\documentclass{aa}  

\usepackage{graphicx}
\usepackage{float}
\usepackage{placeins}
\usepackage{txfonts}

\begin{document}

   \title{Dynamically favorable hosts for submoons around Jupiter and Saturn}

   \author{R.Dahoumane
          \inst{1,2}
          \and
         V. Lainey\inst{2}
         \and
         K. Baillié\inst{2}
          }

   \institute{
   \inst{1}Centro Interdipartimentale di Ricerca Industriale Aerospaziale, Alma Mater Studiorum, Università di Bologna, Forlì (FC), 47121, Italy\\
   \inst{2}LTE, Observatoire de Paris, Université PSL, Sorbonne Université, Université de Lille, LNE,CNRS, 61 Avenue de l'Observatoire, 75014 Paris, France\\
             \email{ryan.dahoumane@unibo.it
             }
             }

   \date{Received XXX; accepted XXX}

  \abstract
   {The long-term stability of small bodies orbiting natural satellites (submoons) in the Solar System has been previously investigated using an analytical approximation of a three-body problem, focusing mainly on tidal forces and neglecting other perturbations. This approach suggests that submoons around most of the regular satellites in the Solar System could be stable. However, no such object has been detected to date.}
   {In this work, we extend these previous results by numerically testing the survival of submoons around a large number of satellites of Jupiter and Saturn. Our goal is to rank the satellites of Jupiter and Saturn according to how dynamically favorable they are for submoon survival over the simulated timescales.}
   {For each regular or irregular satellite considered, we performed 130 separate N-body integrations, each containing one hypothetical submoon with a different initial semimajor axis and inclination. From the 1 Myr survival fractions, we rank the satellites according to their dynamical favorability for submoon survival over the simulated timescale.}
   {We find that, after 1 Myr, the satellites with the highest survival fractions (above 30\%) are Iapetus ($\approx 37\%$), Titan ($\approx 35\%$), Rhea ($\approx 32\%$), Phoebe ($30\%$), Callisto ($\approx 34\%$), and Ganymede ($30\%$).}
   {Rather than providing absolute probabilities for the existence of submoons, these survival fractions identify the most dynamically favorable hosts. Iapetus, Titan, Rhea, Phoebe, Callisto, and Ganymede are the most dynamically favorable submoon hosts among the tested satellites.}


   \maketitle

\section{Introduction}

The existence of binary asteroid systems has been established for several decades \citep{BELTON1996185,PETIT1997}, and their frequency is estimated to be about 2\% in the main belt \citep{Merline_2002} and at least 11\% among the trans-Neptunians \citep{Stephenn_2006}. Today, a few hundred binary asteroids have been detected. Several of these systems have large secondary-to-primary size ratios, including Ida-Dactyl \citep{chapman1995} and Didymos-Dimorphos \citep{PRAVEC2006, NAIDU2020}, which demonstrates that small bodies can themselves host satellites. However, until very recently, the possible existence of objects orbiting planetary satellites had not been considered.

In this paper, we use numerical simulations to identify the best candidates for hosting submoons among  the satellites of the giant planets.
In Section 2, we summarize previous studies of submoon stability. In Section 3, we then discuss the analytical stability in a three-body problem. In Section 4, we develop our methodology. In Section 5, we examine the survival of submoons around the satellites of Jupiter and Saturn. Finally, in Section 6, we present our conclusions and perspectives.

\section{Previous works}

 Early studies, such as \cite{dvorak1986}, \cite{dvorak1989}, \cite{Holman1997}, and \cite{Holman1999} investigated the stability limits of planets in a binary-star system, and laid the groundwork for later three-body stability studies, such as \cite{Barnes_2002} and \cite{Domingos_2006}, which provided the basis for \cite{Kollmeier_2019}. The work of \cite{Kollmeier_2019} was among the first to investigate the potential existence of such "submoons". The authors note that, despite the large number and diversity of satellites orbiting the planets of the Solar System, no submoon has yet been detected. By drawing an analogy between a Star-Planet-Satellite system and a Planet-Satellite-Submoon system, their paper investigates the stability of prospective submoons orbiting the moons of the Solar System's giant planets. 
 
\cite{Kollmeier_2019} adopted the tidal-migration framework developed by \cite{Barnes_2002} and \cite{Sasaki_2012}. These studies, together with \cite{Piro_2018}, showed that tidal friction can cause a satellite either to migrate inward until it collides with its host planet, to migrate outward until it escapes beyond the planet's Hill sphere, or to migrate outward before reversing direction as the planet spins down. Using the orbital stability limit derived by \cite{Holman1999}, \cite{Barnes_2002} then obtained an expression for the maximum lifetime of a satellite undergoing tidal migration around an exoplanet. This expression, which takes into account the tides raised by the satellite on the planet, by the star on the planet, and by the planet on the satellite, is given as a function of the masses of the three objects (Eq. (12) from \cite{Barnes_2002}). From this, \cite{Kollmeier_2019} derived an expression for the maximum mass that allows a submoon to survive for 4.6 Gyr (the age of the Solar System), later corrected by \cite{Rosario-Franco_2020}:

\begin{equation}
R_{\text {moon }} \geq\left[\frac{39 M_{\text {sub }} k_{2, \text { moon }} T \sqrt{G}}{2\left(\frac{4}{3} \pi \rho_{\text {moon }}\right)^{8 / 3} Q_{\text {moon }}}\left(\frac{3 M_p}{(f a_{moon})^3}\right)^{13 / 6}\right]^{1 / 3}\text{,}
\end{equation}

where $M$, $R$, $\rho$, $a$, $Q$ , and $k_2$ denote the mass, radius, density, semimajor axis, tidal quality factor, and tidal Love number of a body, respectively; with subscripts $p$, $moon$, $sub$ referring to the planet, moon, and submoon, respectively. Here, $T$ is the time, $f$ is a stability factor, and $G$ is the gravitational constant.

\cite{Kollmeier_2019} based their calculations on the implicit assumption that all submoon orbits are prograde, which led them to adopt the stability factor $f \approx 0.4895$. However, 
for a coplanar and circular Star-Planet-Satellite system, \cite{Domingos_2006} showed that a retrograde satellite could be stable up to a fraction $f \approx 0.9309$ of the planet's Hill radius. Indeed, the authors derived two different expressions for the prograde and retrograde cases:
\begin{equation}
  f= 0.4895 (1-1.035e_p-0.2738e_{sat})  
\end{equation}
for prograde cases, and
\begin{equation} 
f= 0.9309\ (1-1.0764e_p-0.9812e_{sat}) 
\end{equation}
for retrograde cases. Thus, if only the prograde coefficient is considered, this underestimates the stable region available for submoons.

In spite of that, the study showed that, under this framework, many moons in the Solar System (including the Moon, Europa, Callisto, Ganymede, Titan and Iapetus) could host submoons with radii of several kilometers. Since none has been discovered to date, the authors inferred that more complex dynamical mechanisms must be at work to explain the lack of detected submoons.
These values of $f$ were later updated by \cite{Rosario-Franco_2020} and \cite{Quarles_2021} for the prograde and retrograde cases, respectively. As in \cite{Domingos_2006}, these studies performed large suites of $N$-body integrations of hierarchical three-body problems, varying the eccentricities and semimajor axes of the secondary and tertiary bodies, and identifying the regions where the tertiary survived after a given time span ($10^4$ planetary orbits for \cite{Domingos_2006} and $10^5$ yr for \cite{Rosario-Franco_2020} and \cite{Quarles_2021}). However, \cite{Rosario-Franco_2020} and \cite{Quarles_2021} considered multiple mean longitudes for the same combinations of semimajor axes and eccentricities, and considered an orbit stable only if the object survived for all values of the considered mean longitudes. This allowed them to "update" the \cite{Domingos_2006} coefficients more conservatively. For prograde orbits, \cite{Rosario-Franco_2020} gave a coefficient equal to 
\begin{equation}
    f=0.4061 \ (1-1.1230e_p-0.1862e_{sat})
\end{equation} 
for satellites, and
\begin{equation}
    f= 0.3210 \ (1-0.2757e_p-1.0687e_{sat})
\end{equation}
for prograde submoons. Concerning retrograde satellites, \cite{Quarles_2021} found 
\begin{equation}
    f=0.6684 \ (1-1.236e_p)
.\end{equation}

\cite{Patel_2025} also examined the possibility of submoons existing in exoplanetary systems. Using an approach similar to that of \cite{Rosario-Franco_2020}, the authors studied HD~23079, Kepler-1625, and Kepler-1708. They varied the semimajor axis and inclination of the submoons, with inclinations ranging from 0 to $60^\circ$ relative to the moon's orbital plane. For these exoplanetary configurations, they showed that submoons are not stable for inclinations $>40^\circ$, and that no submoon survived $10^5$~yr for inclinations $>50^\circ$.

The exoplanetary systems studied by \cite{Rosario-Franco_2020} and \cite{Patel_2025} are systems in which the existence of exomoons is strongly suspected due to variations in the transit durations of exoplanets supporting this hypothesis \citep{teachy_2018}. These methods favor the detection of very large exomoons, with a radius similar to that of Neptune in the case of Kepler-1625.
Consequently, the specific systems and numerical configurations considered in \cite{Rosario-Franco_2020} and \cite{Patel_2025}, which involve very massive bodies and planets close to their host stars, differ substantially from the Jovian and Saturnian systems, whose satellites are generally smaller and whose planets orbit farther from the Sun. Although the dynamical mechanisms and general stability structures identified in these studies remain relevant to any submoon problem, their quantitative results therefore cannot be transferred directly to the Solar System.

In the Solar System context, \cite{Sucerquia2024} is also worth mentioning. First, the authors used the equations given in \cite{Domingos_2006} to assess whether a particle could remain in orbit around satellites. In particular, Figure (2) from \cite{Sucerquia2024}  plots the ratio of the Hill radius to the Roche radius for all the large moons of the Solar System (those with mass $>10^{19}$~kg).
Using this first method, very similar to \cite{Kollmeier_2019} and ignoring perturbations, the vast majority of the major moons of the gas giants are predicted to have stable submoons. In a second step, the authors then performed a more detailed dynamical analysis using \mbox{$\mathcal{N}$-body} simulations of massless particles initially distributed in the orbital plane of large satellites of the Solar System. Their simulations demonstrate that ring structures should be stable around the majority of the considered satellites (mass $>10^{19}$~kg). Using a similar approach, we aim to rank the best candidates for hosting a submoon among Jupiter's and Saturn's satellites.

\section{Three-body stability}\label{paragraph:3bodystability}

\begin{figure*}[htbp]
  \centering
  \includegraphics[width=1\textwidth]{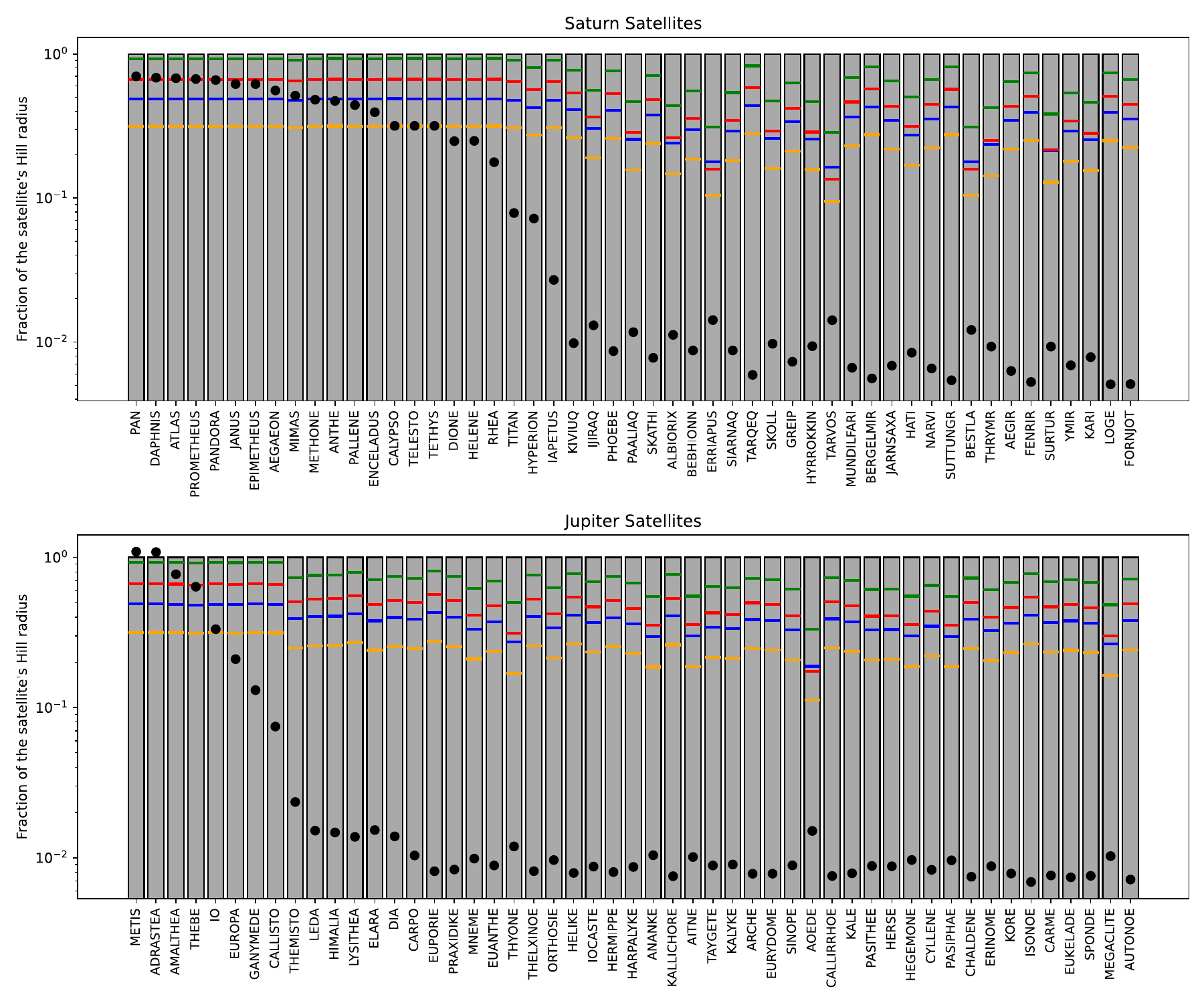}
  \caption{Stability area of satellites of Jupiter and Saturn as a fraction of the satellite's Hill radius. The gray band represents the Hill sphere, the green, red, blue, and yellow bands represent the stability limit for submoons around the satellite in the retrograde case as given by \cite{Domingos_2006}, the retrograde limit as given by \cite{Quarles_2021}, the prograde limit as given by \cite{Domingos_2006}, and the prograde limit as given by \cite{Rosario-Franco_2020}, respectively. The black dot represents the satellite's rigid Roche limit.}
  \label{fig:sucerquia}
\end{figure*}

Using the stability criteria given by \cite{Domingos_2006}, \cite{Rosario-Franco_2020} and \cite{Quarles_2021}, in a similar way to what \cite{Sucerquia2024} did, we can identify which satellites of Jupiter and Saturn could host a submoon, based on the distance between its Roche limit and its stability threshold.
The Hill sphere is defined as the critical distance allowing a massless particle to orbit a secondary body that itself orbits a primary. It is commonly defined as \citep{HamiltonBurns1992,Murray_Dermott_2000}:

\begin{equation}
    R_H = a_1\left(1-e_1\right)\sqrt[3]{\frac{m_1}{m_0}} \text{ ,}
\end{equation}with $R_H$ the Hill radius of the secondary, $a_1$, $e_1$ , and $m_1$  the semimajor axis, eccentricity, and mass of the secondary, respectively, and $m_0$ the mass of the primary.
However, \cite{Domingos_2006} showed that the outer stability limit could be defined more precisely by distinguishing the prograde and retrograde motion of the particle. We can redefine the stability limit of a particle in orbit around a secondary as 
\begin{equation}
    a_E = f \space R_H \text{ ,}
\end{equation}with $f$ a dimensionless coefficient describing the maximum fraction of the Hill radius up to which a particle can be stable around its host. As explained in the previous section, \cite{Domingos_2006}, \cite{Rosario-Franco_2020}, and \cite{Quarles_2021} found different values for $f$ depending on the eccentricities of the bodies in the problem, and on whether the orbits are prograde or retrograde.
By definition, a third body orbiting the secondary can only be stable if the Roche limit of the secondary lies inside its outer stability boundary. 
As in \cite{Sucerquia2024}, by computing the rigid Roche limit of each satellite and comparing it to its stability limit, we can obtain a first overview of the list of satellites expected to be able to host submoons. The rigid Roche limit of a satellite is defined as \citep{Aggarwal_1974,Sucerquia2024}:
\begin{equation}
    r_R = R_{sat}\left(2\frac{\rho_{sat}}{\rho_{sub}}\right)^{1/3}
,\end{equation}with $R_{sat}$ the radius of the satellite, and $\rho_{sat}$ and $\rho_{sub}$ the densities of the satellite and submoon, respectively. Figure \ref{fig:sucerquia} compares the stability limits for submoons predicted by the previously cited models with the host satellite's Roche limit for many Saturnian and Jovian satellites. If the Roche radius lies between the prograde and retrograde stability limits, then only retrograde submoons could be hosted around the satellite. If the Roche radius is greater than the retrograde stability limit, the stable region is effectively empty and no submoon should be able to survive around that satellite. Based on this simple criterion, we see on Figure \ref{fig:sucerquia} that most of Jupiter's and Saturn's satellites should be able to host submoons, with the exception of Jupiter's two innermost moons (Metis and Adrastea).
Figure \ref{fig:sucerquia} can be compared to Figure 2 of \cite{Sucerquia2024} but we added most of Jupiter's and Saturn's satellites, as well as the impact of the satellite's eccentricities on their submoon stability limit. The values for the semimajor axes and eccentricities of each moon are taken from the ephemeris shown in Table \ref{tab:ephemerides}. The mass and radius of the Sun and planets were extracted from the SPICE Kernel \cite{ACTON199665}, while the satellites' masses and radii are available in the Appendix. 
However, this method cannot be applied as such to most satellites of the Solar System, because it assumes a coplanar three-body problem while the vast majority of the satellites in the Solar System are not coplanar with their host planet's orbital plane. To obtain a more realistic survival rate around these satellites, we must conduct a large number of numerical integrations of the $\mathcal{N}$-body problem.

\section{Methodology}\label{sect:methodo}
\subsection{Physics modeling}
To map submoon survival around the satellites of Jupiter and Saturn, we carried out $\mathcal{N}$-body simulations of hypothetical submoons orbiting these satellites. For each host satellite, we ran 130 simulations, each containing one hypothetical submoon with a different initial condition. Each simulation was integrated for 1 Myr, after which we determined the number of cases in which the submoon had neither been ejected nor had it collided with its host satellite.
Each hypothetical submoon has a density of $1\space g/cm^3$ and a radius of $100$~m. The orbital initial conditions of the submoons are available in Table \ref{tab:eo_ss}. The inclination was sampled at 13 linearly spaced values, as in \cite{Holman1997}, while the semimajor axis was sampled at 10 values. The initial positions and velocities of the satellites were taken from the Laboratoire Temps Espace (LTE; formerly the Institut de mécanique céleste et de calcul des
éphémérides, IMCCE; see Table~\ref{tab:ephemerides}). The ephemeris of the Sun, Jupiter, and Saturn were based on \cite{inpop21a_ephe}, and the sources of the masses and radii used for each object are provided in the Appendix. All ephemeris were taken at 12PM on 1 January 2000.

Each simulation included the submoon, the host satellite, Jupiter, Saturn, and the Sun. Additionally, Titan was included in all simulations involving Saturnian moons, as it is Saturn's most massive satellite, and is therefore the most likely to perturb submoons orbiting nearby satellites. Io was included in simulations of submoons around Amalthea and Thebe, while all four Galilean satellites were included in simulations of submoons around any Galilean moon. We also included the $J_2$ term of the host planet, although the effect of the planet's oblateness should, by definition, be negligible for irregular satellites. Indeed, \cite{Nesvorný_2003} define the irregular satellites as those whose orbital plane precession is mainly controlled by the Sun rather than by the $J_2$ of the host planet. They give the semimajor axis above which the Sun's influence on the orbital plane precession becomes greater than that of the planet's oblateness as $a_{irreg} \approx\left(2m_p J_2 R_p^2 a_p^3/m_\star\right)^{1/5}$ with $m_p$, $m_\star$, $R_p$ , and $a_p$ the mass of the planet, the mass of the Sun, the planet's equatorial radius, and the planet's semimajor axis, respectively.
Using values from Table \ref{tab:airreg_constants}, we find that the influence of Jupiter's $J_2$ becomes weaker than the solar perturbation above $\approx 2.32\times10^6$~km (between the orbits of Callisto and Themisto, whose semimajor axes are about $1.88\times10^6$~km and $7.51\times10^6$~km, respectively). For Saturn, the same computation gives a limit of $\approx2.5\times10^6$ km (between the orbits of Hyperion and Iapetus, whose semimajor axes are about $1.50\times10^6$~km and $3.56\times10^6$~km, respectively) beyond which the solar gravitational perturbation dominates over the planet's oblateness. 

\begin{table}[t]
\caption{Physical parameters of Jupiter and Saturn. }
\label{tab:airreg_constants}
\centering
\begin{tabular}{lccccc}
\hline\hline
Body & $m$ (kg) & $R_{\mathrm{eq}}$ (km) & $J_2$  \\
\hline
Jupiter & $1.898\times10^{27}$   & 71492 &  0.014696  \\
Saturn  & $5.683\times10^{26}$   & 60330 &  0.016291  \\
\hline
\end{tabular}
\tablefoot{Values come from \cite{Tremaine_2009}}
\\[1mm]
\raggedright
\end{table}

When included, the spin axis of the host planet is assumed to be fixed throughout each simulation. We modeled all the satellites as point masses. In reality, irregular satellites often deviate significantly from sphericity \citep{THOMAS_1989}. This non-sphericity may affect the stability of submoons; however, their spin axes are themselves unstable on million-year timescales, so we could not properly include the shape effects without significantly complicating the problem by considering many different spin-axis configurations.  

For satellites on inclined orbits, the submoon orbital elements were generated in a local frame whose reference plane is the osculating orbital plane of the host satellite around the planet. The resulting Cartesian state vectors were then transformed back into the global integration frame.

\subsection{Numerical integrations}
\subsubsection{Solving $\mathcal{N}$-body problems}
To solve these $\mathcal{N}$-body problems, we used our own code, which was previously used in \cite{Dahoumane_2025}. We used a fixed-axis satellite-centric frame centered on the host satellite. This choice improves the numerical precision of the relative position and velocity of the submoon with respect to its host satellite. The expressions for the accelerations in this frame are given in Appendix \ref{app:force}.

We integrated these equations with the Gauss-Radau integrator described in \cite{Everhart_1985}, with the LL coefficient set to 13. We chose to use the original Gauss-Radau integrator rather than the IAS15 \citep{REINIAS15}, as the latter is significantly slower than the former. This higher computational cost is accompanied by greater numerical accuracy. To reduce numerical errors in the relative positions and velocities obtained with our Gauss-Radau implementation at LL=13, which could otherwise affect the detection of collisions or ejections, we placed the origin of the coordinate system at the center of the host satellite. We performed a convergence test on Iapetus, one of the key long-lived cases identified below, and found that increasing the accuracy from LL13 to LL14 does not significantly change the survival fraction (see Appendix).

\begin{table}[!ht]
\caption{Initial orbital elements of submoons around each satellite. }
    \centering
    \begin{tabular}{ccc}
    \hline
        Orbital element & Value & Number of samples \\ \hline
        a & [$r_{Roche}$, $0.9309 r_{Hill}$] & 10 \\ 
        e & 0 & 1 \\ 
        i & [0°, 180°] & 13 \\  \hline
    \end{tabular}
    \tablefoot{For each host satellite, we performed 130 separate simulations, each containing one hypothetical submoon. The 130 initial conditions combine ten semimajor-axis values and 13 inclination values. The inclination is given with respect to the satellite's orbit around the planet. Each orbital element is linearly distributed in the given interval (values for the inclination are 0°, 15°, 30° ... 180°). The semimajor axis is distributed within the Roche limit and 0.9309 times the Hill sphere of the satellite. The argument of periapsis, longitude of the ascending node, and mean anomaly are taken as equal to 0.}
    \label{tab:eo_ss}
\end{table}

\subsubsection{Collision detection}\label{subsect:colldetect}
Reliable collision detection is required for long-term integrations. A simple approach to collision detection in an \mbox{$\mathcal{N}$-body} simulation is to monitor the separation 
$d$ between the two bodies at each time step:
\begin{equation}\label{eqn:coll_cond}
d \leq R_i +R_j \implies \mathit{collision}
.\end{equation}This method is simple but Appendix \ref{app:detection} shows that, when used with our implementation of the Gauss-Radau integrator, it performs adequately for the relatively simple configurations considered here (collisions between a secondary body on an elliptical trajectory around a primary).

\subsubsection{Ejection detection}\label{subsect:eject}

To avoid unnecessarily continuing the integration after a submoon becomes unbound from its host satellite, we introduced an operational ejection criterion. When the criterion is met, the integration stops.

To do so, we let $m_0$ be the mass of the central body and $m_1$ the mass of a secondary orbiting the central body (in our case, they are respectively the satellite and submoon). 
In the absence of dissipation and neglecting the spin of the objects, the total energy $E_{tot}$ is simply

\begin{equation}
    E_{tot} = E_{k}+E_{p}
,\end{equation}where $E_k$ and $E_p$ are the kinetic and potential energies, respectively.

In the unperturbed two-body problem, the sign of the orbital energy distinguishes bound and unbound Keplerian trajectories:

\begin{equation}
    \label{Eq: ec_ep}
    \left\{
        \begin{array}{ll}
            E_{tot}  < 0 \implies Ellipse\\
            E_{tot} >0 \implies Hyperbola
        \end{array}
    \right.
\end{equation}In a frame centered on the body $0$, if all other bodies are neglected, we have (equations (22) and (23) from \cite{Lainey2004}):
\begin{equation}
\label{Eq:Ec_Ep_inertial}
    \left\{
        \begin{array}{ll}
          \displaystyle  E_k  = \frac{m_0m_1 \mathbf{V}_1^2}{2(m_0+m_1)}\\
          \displaystyle  E_p  = \frac{-Gm_0m_1}{r_1} 
        \end{array}
    \right.
.\end{equation}Here, $\mathbf{r}_1$ and $\mathbf{V}_1$ denote the position and velocity vectors of the submoon relative to the host satellite, respectively, and $r_1=\lVert\mathbf{r}_1\rVert$. We therefore consider a submoon to be ejected when its instantaneous two-body energy relative to the host satellite becomes positive, and stop the integration at that point. In the perturbed multi-body system considered here, this condition is used as a practical stopping criterion rather than as a rigorous proof of permanent escape. This energy-based operational criterion is similar to that adopted by \cite{Domingos_2006}, following \cite{Neto_Winter_Melo_2004}.

\section{Submoon survival around Jupiter and Saturn}\label{sect:gasgiant}
\subsection{Survivability after 1 Myr}

\begin{figure*}[htbp]
\sidecaption
\centering
  \includegraphics[width=0.7\textwidth]{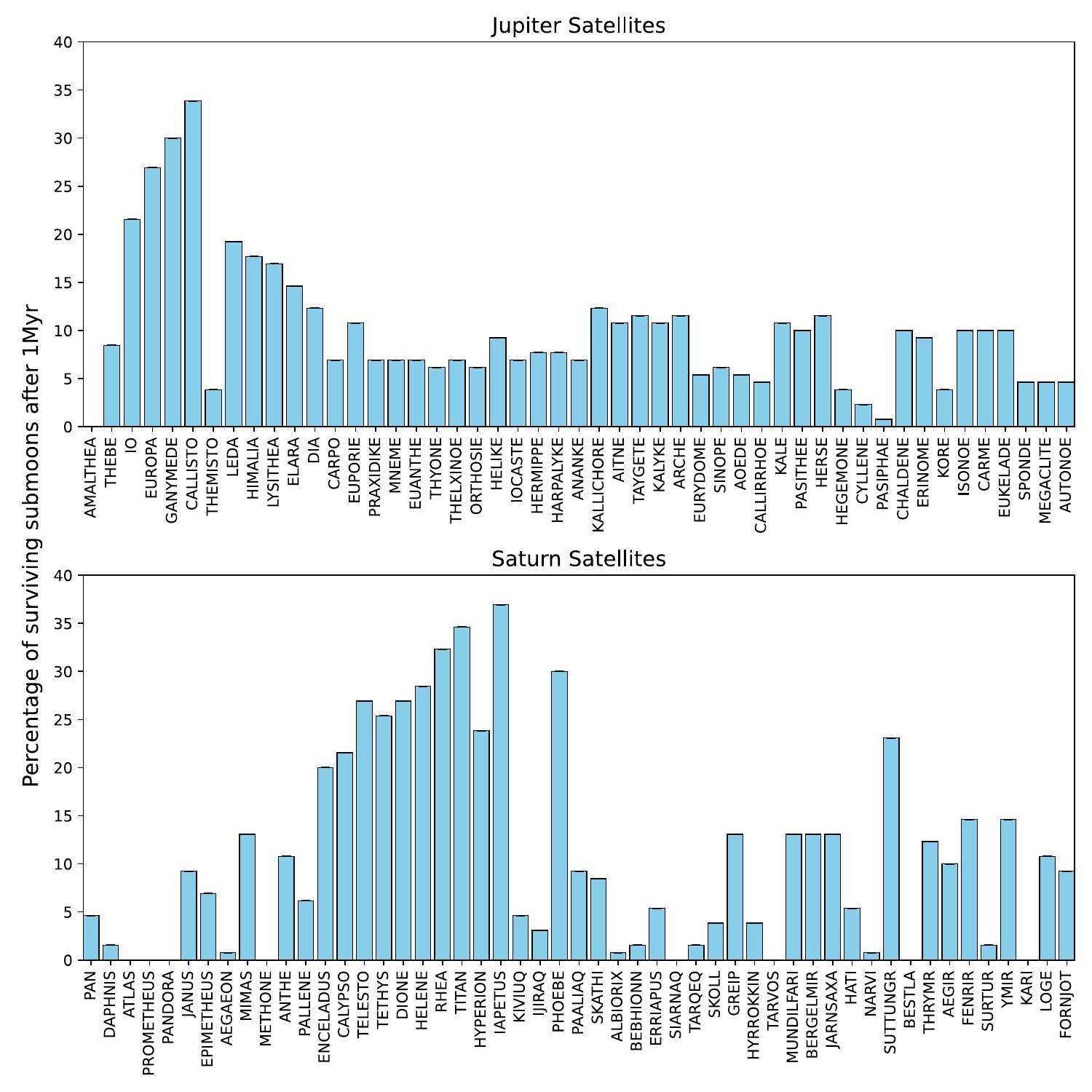}
  \caption{Detailed survival percentages of submoons around actual satellites of Jupiter and Saturn. Each bar shows the fraction surviving after 1 Myr among the 130 separate initial conditions tested for one host satellite.}
  \label{fig:survie_sat_vrai}
\end{figure*}

\begin{figure}
    \centering
  \includegraphics[width=0.49\textwidth]{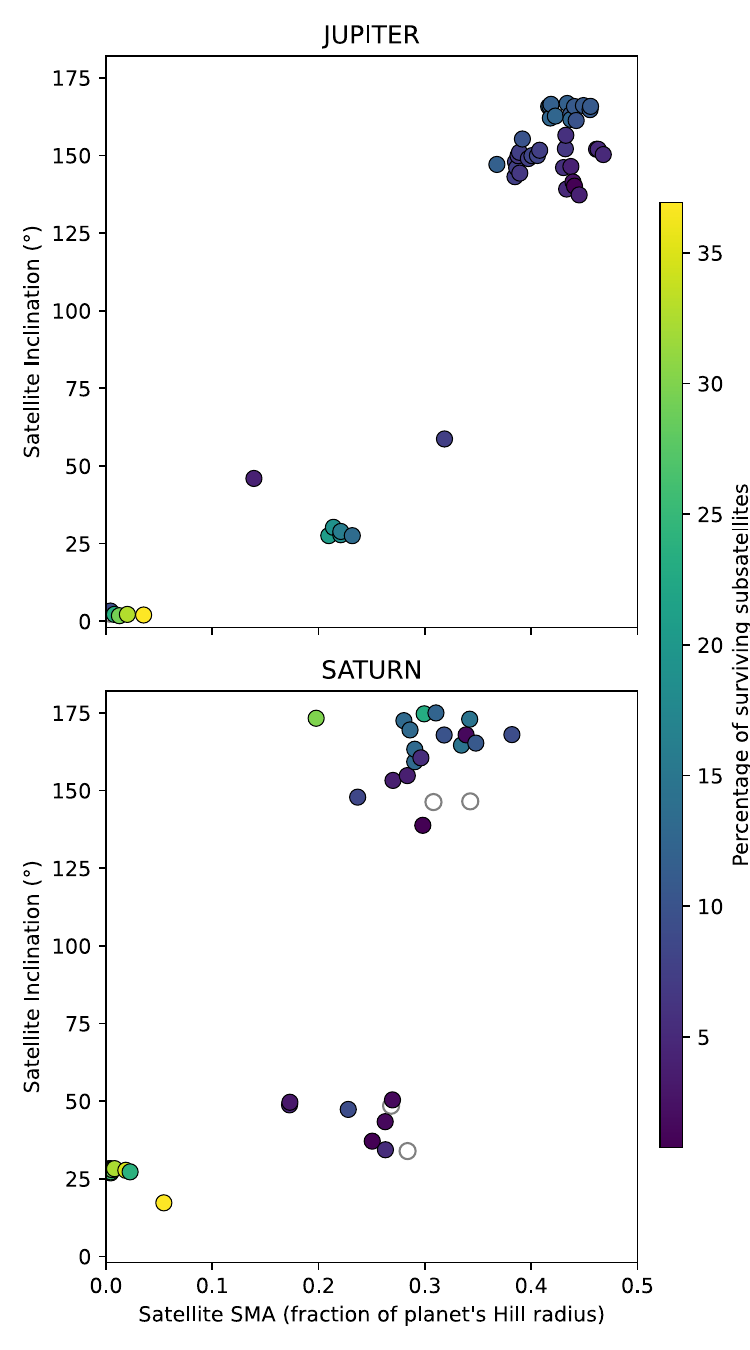}
  \caption{Survivability of submoons around actual satellites of Jupiter and Saturn. Each dot represents an existing Jovian or Saturnian satellite, and its color represents the survival percentage of the 130 submoons that were initially orbiting it, after 1 Myr. White dots represent a 0\% submoon survival rate. The submoon initial conditions are detailed in Section \ref{sect:methodo} and Table \ref{tab:eo_ss}. These results are identical to Figure \ref{fig:survie_sat_vrai}, but are displayed here to highlight their distribution in the orbital-parameter space of the host satellites.}
  \label{fig:survie_sat_vrai2}
\end{figure}

\begin{table*}[!ht]
    \caption{Sources of the satellite ephemeris used to compute the survivability around satellites of giant planets.}
    \centering
    \begin{tabular}{cc}
    \hline
        Moons  & Ephemeris \\ \hline
        Io, Europa, Ganymede, Callisto, Amalthea, Thebe , Adrastea & \cite{noe5-2021_ephe} \\ 
        Irregular Jovian satellites & \cite{sai5-2021_ephe} \\ 
        Mimas, Enceladus, Tethys, Dione, Rhea, Titan, Hyperion, Iapetus & \cite{noe6-2018_ephe} \\ 
        Phoebe & \cite{phoebe2020_ephe} \\ 
        Other Saturnian moons & \cite{sai6-2021_ephe} \\ \hline
    \end{tabular}

    \label{tab:ephemerides}
\end{table*}

\subsubsection{Best candidates around Jupiter and Saturn}
Figure \ref{fig:survie_sat_vrai} shows that, among the Jovian moons, those that exhibit a significant submoon survival fraction (exceeding 25\%) are the Galilean satellites Europa, Ganymede and Callisto, which is consistent with the findings of \cite{Sucerquia2024}. Among them, only Ganymede and Callisto reach a survival fraction of 30\%. Such high survival fractions are mainly explained by their large Hill sphere, combined with their low eccentricities and low inclinations. Furthermore, the large eccentricities of irregular satellites are additional factors that tend to destabilize submoons \citep{Domingos_2006, Rosario-Franco_2020, Quarles_2021}, thus lowering the survival rate of these objects.

The decrease in the survival rate that accompanies an increase in inclination is characteristic of the Lidov-Kozai mechanism \citep{LIDOV1961,Kozai_1962}. The secular perturbations caused by the presence of the planet (and to a lesser extent the star) on the submoon's orbit tend to create exchange cycles between the submoon's inclination and eccentricity. A high inclination of the submoon's orbit around the moon relative to the moon's orbit around the planet can lead, over the course of secular cycles, to a sharp increase in eccentricity accompanied by a decrease in inclination. It is therefore expected that after 1 million years, the number of surviving submoons decreases as the initial inclination increases. Figure \ref{fig:survie_sat_vrai2} also suggests a broader dependence on the orbital configuration of the host satellite: satellites on more distant and more inclined orbits generally exhibit lower submoon survival fractions. This trend is qualitatively consistent with the dependence of hierarchical-system stability on orbital separation and mutual inclination described by \cite{10.1093/mnras/stw3096}.

In contrast with the Jovian system, eight satellites exceed the 25\% survival rate among Saturn's moons: Telesto, Tethys, Dione, Helene, Rhea, Titan, Iapetus, and Phoebe. Among these eight moons, four of them reach a 30\% survival rate: Rhea, Titan, Iapetus, and Phoebe. Titan is the most massive satellite of Saturn, while still having an orbit close to its parent planet. Moreover, other satellites close to Saturn have their submoons' orbit perturbed by Titan, in addition to Saturn and the Sun, while the main perturbation exerted on Titan's submoons is exclusively due to Saturn. On the other hand, Phoebe's relatively high survival rate can likely be attributed to its mass — the largest among the retrograde satellites considered — and its proximity to Saturn. Its retrograde, nearly planar orbit and relatively low eccentricity further enhance its stability.

Interestingly, among all the tested satellites, Iapetus has the highest submoon survival rate after 1 Myr, making it the most dynamically favorable host for a submoon among the well-known satellites of Jupiter and Saturn. Although tidal effects were not included in our simulations, they are weak for submoons of the size considered here (see Appendix \ref{app:nonconservative}). Such a high short-term survival rate is nevertheless consistent with the scenario proposed by \cite{LEVISON2011773} and \cite{Dombard_2012}, in which a former submoon of Iapetus ultimately re-impacted the satellite and contributed to the formation of its equatorial ridge.

We also notice that, in our simulations, in spite of being very close to Saturn, Pan and Daphnis have a couple of surviving submoons in their neighborhood. This is quite surprising for such inner moons; however, in reality, the perturbation due to Saturn's rings would probably destabilize this small remaining fraction of submoons.

\subsubsection{Submoons around the Moon}
\begin{figure}[htbp]
  \centering

  \includegraphics[width=\linewidth]{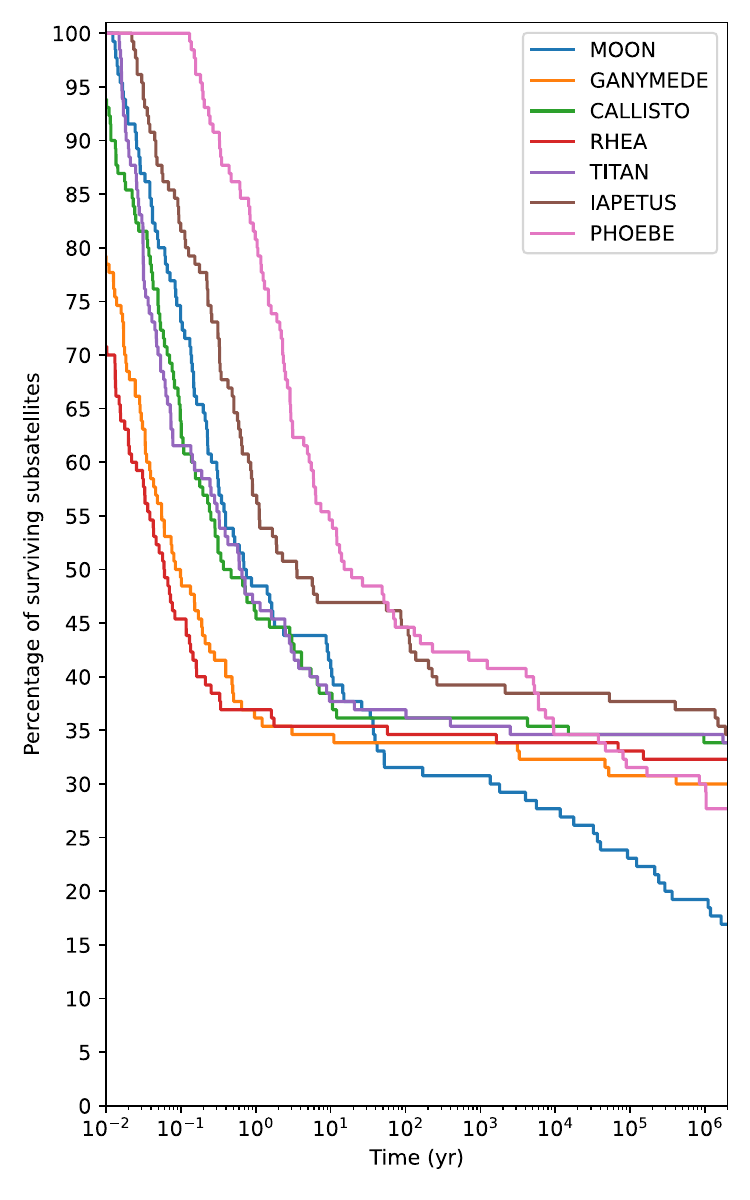}
  \caption{Evolution of the submoon survival fractions up to 2 Myr for the satellites whose survival fraction at 1 Myr was at least 30\%, together with the Moon. The fact that some moons start losing their submoons later than others is simply due to differences in their orbital periods. All satellites start to lose submoons after approximately $10^{-1}$ orbital period.}
  \label{fig:long_terme}
\end{figure}

To compare these results with a better-known case, we also applied the same procedure to the Moon, around which no stable submoon has ever been detected. We therefore conducted a large number of Sun-Earth-Moon-Submoon simulations in the same way as we did for Saturn's and Jupiter's satellites. Our results (see Figure \ref{fig:long_terme}) show that the Moon has a submoon survival fraction of approximately 19\%, compared with a median survival fraction of approximately 8\% among the Jovian and Saturnian satellites considered. For comparison, Figure \ref{fig:survie_sat_vrai} shows that, among Jupiter's satellites, only Leda and the four Galilean moons have a survival rate above 19\%. Among Saturn's satellites, Enceladus, Calypso, Telesto, Tethys, Dione, Helene, Rhea, Titan, Hyperion, Iapetus, Phoebe, and Suttungr have a survival rate above that of the Moon.

\subsection{Survival between 1 and 2 Myr}

After obtaining such high survival rates for the satellites mentioned above, we now need to test whether these survival fractions still rapidly evolve beyond 1 Myr.
To do this, we selected the satellites with the highest submoon survival rates, as well as the Moon, and extended the integrations to 2 Myr. We were not able to extend our simulations any longer due to our demanding physical modeling.

Figure \ref{fig:long_terme} shows that most unstable submoons are lost during the early phase of the simulations. After 1 Myr, the survival curves no longer show a rapid decay, although a limited number of additional losses still occur for some satellites. No additional submoon is lost between 1 and 2 Myr for Ganymede, Callisto, or Rhea. By contrast, Titan loses one additional submoon, while Iapetus, Phoebe, and the Moon continue to lose a few submoons during the second million years. Since each satellite was tested with 130 initial submoon configurations, these late losses correspond to variations of only a few percent in survival fractions. This suggests that the 1 Myr survival fractions are not measured during a rapid ongoing decay, but rather after the removal of the most unstable initial configurations.
Consequently, doubling the integration time does not fundamentally change the ranking inferred from the 1 Myr simulations, although the survival fractions of some candidates, especially Phoebe and Iapetus, decrease slightly. Figure \ref{fig:1_5myr} in the Appendix shows that the remaining submoons after 2 Myr are those close to the satellite's orbital plane, with a relatively low semimajor axis. The submoons lost after the first million years of integration are those with the highest semimajor axis and furthest from the orbital plane.

However, on longer timescales, effects such as the Binary Yarkovsky–O'Keefe–Radzievskii–Paddack (BYORP) and Yarkovsky effects could potentially modify the semimajor axes of the submoons by hundreds of kilometers (see Appendix). This does not necessarily imply that the submoon would be ejected or collide with its host satellite, as the Hill spheres of the satellites with the most surviving submoons are typically a few thousand kilometers, but without a deeper study, we cannot state with confidence that they would not reduce the survival after tens of millions of years. The domination of BYORP or Binary Yarkovsky depends on the rotation state of the secondary (close to synchronicity or not), so the integration of rotation would be required \citep{Zhou_2024}. The study of such forces is beyond the scope of the present study, as rotation of the submoons was not modeled here. Although an integration of satellite rotation with their orbital motion was performed in \cite{Dahoumane_2025}, the authors did it on a 25,000 years timescale, which is far shorter than the million years considered here.

\section{Conclusions and discussions}
This study identifies Iapetus, Titan, Rhea, Phoebe, Callisto, and Ganymede as those satellites that provide the most dynamically favorable conditions for long-lived submoon survival under the assumptions of our model.
These simulations were performed under the assumption that all bodies apart from planets behave as point masses; consequently, we neglected the shape of the satellites, and assumed circular initial orbits for the submoons.
 However, in this study, we did not aim to estimate the exact stability of submoons around each satellite, but rather to rank the most dynamically favorable submoon hosts among the well-known satellites of Jupiter and Saturn.
A natural next step would be to include chaos indicators in order to characterize the stable regions around those satellites more precisely. 

Although our simulations identify the satellites that are the most dynamically favorable for submoon survival under the adopted model, they do not address whether the simulated 100~m objects would be detectable. Consequently, this dynamical ranking cannot be interpreted as a list of observational priorities without further investigation.
Nevertheless, several observational techniques could be investigated in future dedicated detectability studies.
Stellar occultation campaigns have been conducted to study the natural satellites of planets; these have led to a significant improvement in our understanding of their properties (astrometry, shape, rotation) \citep{gomes_2020}. This technique has also been used to detect and characterize satellites around small bodies, such as asteroid 4337 Arecibo \citep{Gault2022Arecibo, Liu_2025, Lallemand2026}. Automated methods applied to spacecraft images may provide another avenue for searching large image archives \citep{Quaglia_2025}. Determining whether these approaches could detect submoons of the sizes and at the orbital separations considered here would require a dedicated quantitative analysis that is out of the scope of this study.

The simulations we performed consider very small submoons (100~m in radius) on which tidal forces are therefore negligible over the timescales considered here. However, it is possible that, in the past, certain satellites of the Solar System had more massive submoons that migrated under tidal effects (or via the Binary Yarkovsky effect or BYORP \citep{Zhou_2024}), thereby being ejected or fragmenting after passing below the Roche limit. This is notably the scenario proposed by \cite{LEVISON2011773} to explain the equatorial ridge of Iapetus, and the high survival rate of submoons around Iapetus may be considered consistent with such a scenario.

\section*{Data availability}
The files containing the initial conditions for each of the simulations presented in the present article are available at https://doi.org/10.5281/zenodo.19820511

\begin{acknowledgements}
      The authors would like to thank Tilmann Denk and Scott Sheppard for their work on outer moons of Jupiter and Saturn, and making estimations of outer satellites properties available on their Websites. The authors would also like to thank the anonymous reviewer for their feedback which allowed us to greatly improve the quality of this paper.
\end{acknowledgements}

\bibliographystyle{aa}
\bibliography{sources}
\begin{appendix}
\raggedbottom
    \section{Numerical robustness tests}
    \subsection{Integrator robustness}
        \begin{figure}[!htbp]
  \centering
  \includegraphics[width=\linewidth]{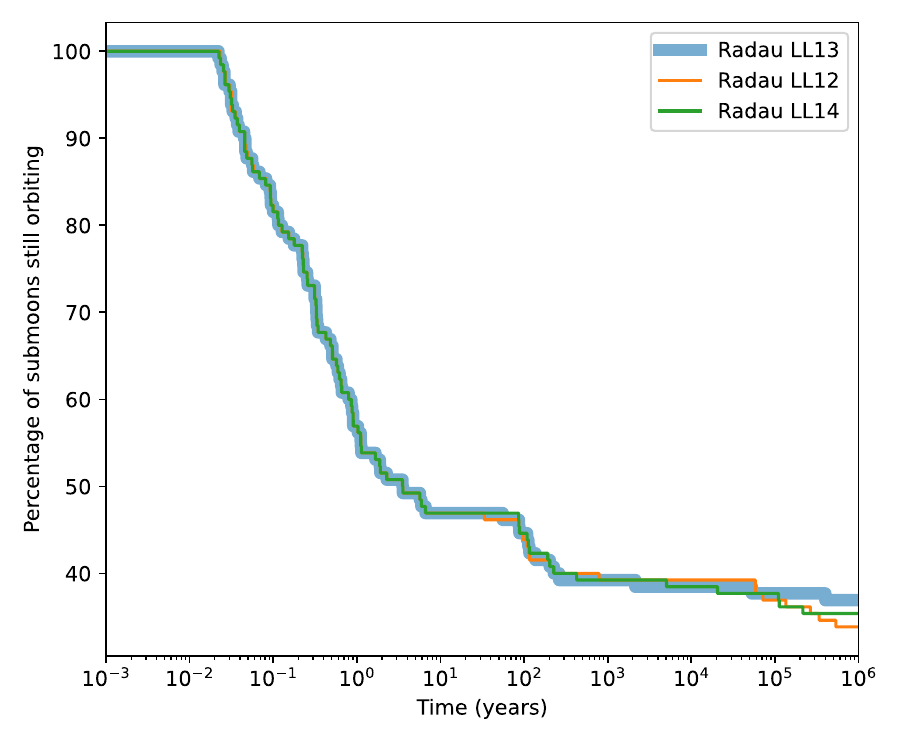}
  \caption{Submoon survival fraction around Iapetus as a function of time, for three different values of the precision parameter LL of the Gauss-Radau integrator \citep{Everhart_1985}}
  \label{fig:compare_radau}
\end{figure}
    To assess the numerical robustness of the million-year integrations, we repeated the full set of 130 Iapetus initial conditions with the Gauss-Radau integrator \citep{Everhart_1985} using three different values of the accuracy parameter, LL=12, LL=13, and LL=14. Iapetus was chosen as a representative long-lived case because it is one of the main candidates identified in this study. The survival curves remain very similar over the full integration time. Figure \ref{fig:compare_radau} shows that at 1 Myr, increasing the accuracy from our nominal value LL=13 to LL=14 changes the final survival fraction by less than $2$ percentage points, corresponding to 2 trajectories out of 130. We therefore do not claim convergence at the level of individual chaotic trajectories. However, the ensemble survival fraction is stable within the resolution of our initial-condition sampling. This uncertainty does not affect the identification of the best candidate satellites.

    \subsection{Collision and ejection detection}\label{app:detection}

\begin{figure}[H]
      \centering
      \includegraphics[width=\linewidth]{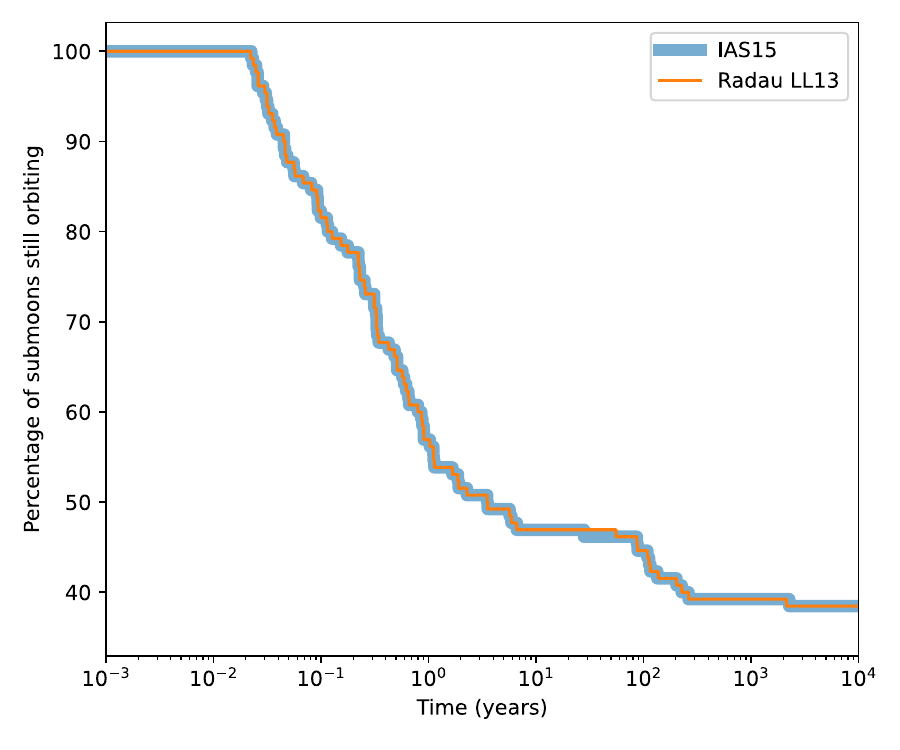}
      \caption{Submoon survival fraction around Iapetus as a function of time, using the IAS15 \citep{REINIAS15} and the Gauss-Radau integrator \citep{Everhart_1985} with the LL parameter set to 13}
      \label{fig:compare_radau_ias}
    \end{figure}

    The collision criterion used in this work is based on the mutual distance between the submoon and its host satellite, as explained in Section \ref{subsect:colldetect}. A collision is recorded when this distance becomes smaller than the sum of the physical radii of the two bodies. Since this event detection may, in principle, depend on the time-step sequence used by the integrator, we performed an additional consistency check with IAS15 \citep{REINIAS15}. We reran the Iapetus simulations over the first \(10^4\) yr with IAS15 and compared the resulting collision and ejection detections with those obtained using the Gauss--Radau integrator with LL=13. The IAS15 integrator has the advantage of robustness when handling close approaches and collisions, by automatically reducing the timestep when a variation of acceleration is detected. However, the disadvantage is its greater computational cost, which is why we ran this simulation for only 10 000 years.
    As shown in figure \ref{fig:compare_radau_ias}, over the first 10 000 years of integration, IAS15 and the Gauss-Radau integrator produce almost identical detections of collisions and ejections, using the method presented in Sections \ref{subsect:colldetect} and \ref{subsect:eject}. 

\clearpage
    \section{Masses and radii of each moon}\label{App:mass_radii}
    The following tables contain the values of radii and masses of each satellite that was considered in this paper. All the values of the radii of satellites that were used in this paper come from \cite{Sheppard_JupiterMoons_website} and \cite{Sheppard_SaturnMoons_website}. The source of each satellite mass is written in the table below.
    \begin{table}[!ht]
        \caption{Physical parameters of the Jovian satellites.}
    \centering
        \small                
\setlength{\tabcolsep}{4pt} 
    \begin{tabular}{llll}
        Satellite & Radius (km) & Mass (kg) & Source of the mass \\ \hline
        Metis & 22 & 1.00$\times 10^{17 }$ & \cite{Williams_NSSDCA_JovianFact} \\ 
        Adrastea & 8 & 2.00$\times 10^{16 }$ & \cite{Williams_NSSDCA_JovianFact} \\ 
        Amalthea & 84 & 7.50$\times 10^{18 }$ & \cite{Williams_NSSDCA_JovianFact} \\ 
        Thebe & 49 & 8.00$\times 10^{17 }$ & \cite{Williams_NSSDCA_JovianFact} \\ 
        Io & 1821.5 & 8.93$\times 10^{22 }$ & \cite{Williams_NSSDCA_JovianFact} \\ 
        Europa & 1561 & 4.80$\times 10^{22 }$ & \cite{Williams_NSSDCA_JovianFact} \\ 
        Ganymede & 2631 & 1.48$\times 10^{23 }$ & \cite{Williams_NSSDCA_JovianFact} \\ 
        Callisto & 2410.5 & 1.08$\times 10^{23 }$ & \cite{Williams_NSSDCA_JovianFact} \\ 
        Themisto & 4.5 & 3.82$\times 10^{14 }$ & Assumed density of $1\space g/cm^3$ \\ 
        Leda & 9 & 6.00$\times 10^{15 }$ & \cite{Williams_NSSDCA_JovianFact} \\ 
        Himalia & 80 & 9.50$\times 10^{18 }$ & \cite{Williams_NSSDCA_JovianFact} \\ 
        Ersa & 1.5 & 1.41$\times 10^{13 }$ & Assumed density of $1 \space g/cm^3$ \\ 
        Pandia & 1.5 & 1.41$\times 10^{13 }$ & Assumed density of $1 \space g/cm^3$ \\ 
        Lysithea & 19 & 8.00$\times 10^{16 }$ & \cite{Williams_NSSDCA_JovianFact} \\ 
        Elara & 39 & 8.00$\times 10^{17 }$ & \cite{Williams_NSSDCA_JovianFact} \\ 
        Dia & 2 & 3.35$\times 10^{13 }$ & Assumed density of $1 \space g/cm^3$ \\ 
        Carpo & 1.5 & 1.41$\times 10^{13 }$ & Assumed density of $1 \space g/cm^3$ \\ 
        Valetudo & 0.5 & 5.24$\times 10^{11 }$ & Assumed density of $1 \space g/cm^3$ \\ 
        Euporie & 1 & 4.19$\times 10^{12 }$ & Assumed density of $1 \space g/cm^3$ \\ 
        Euanthe & 1.5 & 1.41$\times 10^{13 }$ & Assumed density of $1 \space g/cm^3$ \\ 
        Orthosie & 1 & 4.19$\times 10^{12 }$ & Assumed density of $1 \space g/cm^3$ \\ 
        Thyone & 2 & 3.35$\times 10^{13 }$ & Assumed density of $1 \space g/cm^3$ \\ 
        Mneme & 1 & 4.19$\times 10^{12 }$ & Assumed density of $1 \space g/cm^3$ \\ 
        Harpalyke & 2 & 3.35$\times 10^{13 }$ & Assumed density of $1 \space g/cm^3$ \\ 
        Hermippe & 2 & 3.35$\times 10^{13 }$ & Assumed density of $1 \space g/cm^3$ \\ 
        Praxidike & 3.5 & 1.80$\times 10^{14 }$ & Assumed density of $1 \space g/cm^3$ \\ 
        Thelxinoe & 1 & 4.19$\times 10^{12 }$ & Assumed density of $1 \space g/cm^3$ \\ 
        Eupheme & 1 & 4.19$\times 10^{12 }$ & Assumed density of $1 \space g/cm^3$ \\ 
        Helike & 2 & 3.35$\times 10^{13 }$ & Assumed density of $1 \space g/cm^3$ \\ 
        Iocaste & 2.5 & 6.54$\times 10^{13 }$ & Assumed density of $1 \space g/cm^3$ \\ 
        Ananke & 14 & 1.15$\times 10^{16 }$ & Assumed density of $1 \space g/cm^3$ \\ 
        Arche & 1.5 & 1.41$\times 10^{13 }$ & Assumed density of $1 \space g/cm^3$ \\ 
        Pasithee & 1 & 4.19$\times 10^{12 }$ & Assumed density of $1 \space g/cm^3$ \\ 
        Herse & 1 & 4.19$\times 10^{12 }$ & Assumed density of $1 \space g/cm^3$ \\ 
        Chaldene & 2 & 3.35$\times 10^{13 }$ & Assumed density of $1 \space g/cm^3$ \\ 
        Kale & 1 & 4.19$\times 10^{12 }$ & Assumed density of $1 \space g/cm^3$ \\ 
        Isonoe & 2 & 3.35$\times 10^{13 }$ & Assumed density of $1 \space g/cm^3$ \\ 
        Aitne & 1.5 & 1.41$\times 10^{13 }$ & Assumed density of $1 \space g/cm^3$ \\ 
        Erinome & 1.5 & 1.41$\times 10^{13 }$ & Assumed density of $1 \space g/cm^3$ \\ 
        Taygete & 2.5 & 6.54$\times 10^{13 }$ & Assumed density of $1 \space g/cm^3$ \\ 
        Carme & 23 & 1.00$\times 10^{17 }$ & \cite{Williams_NSSDCA_JovianFact} \\ 
        Kalyke & 2.5 & 6.54$\times 10^{13 }$ & Assumed density of $1 \space g/cm^3$ \\ 
        Eukelade & 2 & 3.35$\times 10^{13 }$ & Assumed density of $1 \space g/cm^3$ \\ 
        Eirene & 2 & 3.35$\times 10^{13 }$ & Assumed density of $1 \space g/cm^3$ \\ 
        Kallichore & 1 & 4.19$\times 10^{12 }$ & Assumed density of $1 \space g/cm^3$ \\ 
        Philophrosyne & 1 & 4.19$\times 10^{12 }$ & Assumed density of $1 \space g/cm^3$ \\ 
        Eurydome & 1.5 & 1.41$\times 10^{13 }$ & Assumed density of $1 \space g/cm^3$ \\ 
        Autonoe & 2 & 3.35$\times 10^{13 }$ & Assumed density of $1 \space g/cm^3$ \\ 
        Sponde & 1 & 4.19$\times 10^{12 }$ & Assumed density of $1 \space g/cm^3$ \\ 
        Pasiphae & 29 & 3.00$\times 10^{17 }$ & \cite{Williams_NSSDCA_JovianFact} \\ 
        Megaclite & 3 & 1.13$\times 10^{14 }$ & Assumed density of $1 \space g/cm^3$ \\ 
        Sinope & 19 & 8.00$\times 10^{16 }$ & \cite{Williams_NSSDCA_JovianFact} \\ 
        Hegemone & 1.5 & 1.41$\times 10^{13 }$ & Assumed density of $1 \space g/cm^3$ \\ 
        Aoede & 2 & 3.35$\times 10^{13 }$ & Assumed density of $1 \space g/cm^3$ \\ 
        Callirrhoe & 3.5 & 1.80$\times 10^{14 }$ & Assumed density of $1 \space g/cm^3$ \\ 
        Cyllene & 1 & 4.19$\times 10^{12 }$ & Assumed density of $1 \space g/cm^3$ \\ 
        Kore & 1 & 4.19$\times 10^{12 }$ & Assumed density of $1 \space g/cm^3$ \\ \hline
    \end{tabular}
\end{table}

\begin{table}[!ht]
\caption{Physical parameters of the Saturnian satellites.}
    \centering
    \small                
\setlength{\tabcolsep}{4pt} 
    \begin{tabular}{llll}
    
        Satellite & Radius (km) & Mass (kg) & Source of the mass \\ \hline
        Pan & 10 & 5.00$\times10^{15 }$ & \cite{Williams_NSSDCA_SaturnianFact} \\ 
        Daphnis & 3.5 & 1.00$\times10^{14 }$ & \cite{Williams_NSSDCA_SaturnianFact} \\ 
        Atlas & 16 & 7.00$\times10^{15 }$ & \cite{Williams_NSSDCA_SaturnianFact} \\ 
        Prometheus & 50 & 1.60$\times10^{17 }$ & \cite{Williams_NSSDCA_SaturnianFact} \\ 
        Pandora & 42 & 1.40$\times10^{17 }$ & \cite{Williams_NSSDCA_SaturnianFact} \\ 
        Epimetheus & 59.5 & 5.30$\times10^{17 }$ & \cite{Williams_NSSDCA_SaturnianFact} \\ 
        Janus & 89 & 1.90$\times10^{18 }$ & \cite{Williams_NSSDCA_SaturnianFact} \\ 
        Aegaeon & 0.25 & 7.82$\times10^{10 }$ & \cite{THOMAS2020} \\ 
        Mimas & 198.5 & 3.79$\times10^{19 }$ & \cite{Williams_NSSDCA_SaturnianFact} \\ 
        Methone & 1.5 & 3.92$\times10^{12 }$ & \cite{THOMAS2020} \\ 
        Anthe & 0.5 & 1.50$\times10^{12 }$ & \cite{Williams_NSSDCA_SaturnianFact} \\ 
        Pallene & 2 & 1.15$\times10^{13 }$ & \cite{THOMAS2020} \\ 
        Enceladus & 249.5 & 1.08$\times10^{20 }$ & \cite{Williams_NSSDCA_SaturnianFact} \\ 
        Tethys & 530 & 6.18$\times10^{20 }$ & \cite{Williams_NSSDCA_SaturnianFact} \\ 
        Telesto & 12 & 7.00$\times10^{15 }$ & \cite{Williams_NSSDCA_SaturnianFact} \\ 
        Calypso & 9.5 & 4.00$\times10^{15 }$ & \cite{Williams_NSSDCA_SaturnianFact} \\ 
        Dione & 559 & 1.10$\times10^{21 }$ & \cite{Williams_NSSDCA_SaturnianFact} \\ 
        Helene & 16 & 3.00$\times10^{16 }$ & \cite{Williams_NSSDCA_SaturnianFact} \\ 
        Rhea & 764 & 2.31$\times10^{21 }$ & \cite{Williams_NSSDCA_SaturnianFact} \\ 
        Titan & 2575 & 1.35$\times10^{23 }$ & \cite{Williams_NSSDCA_SaturnianFact} \\ 
        Hyperion & 133 & 5.55$\times10^{21 }$ & \cite{Jacobson_2022} \\ 
        Iapetus & 718 & 1.81$\times10^{21 }$ & \cite{Williams_NSSDCA_SaturnianFact} \\ 
        Kiviuq & 8 & 1.00$\times10^{15 }$ & \cite{Denk_OuterMoonsSaturn} \\ 
        Ijiraq & 6 & 5.40$\times10^{14 }$ & \cite{Denk_OuterMoonsSaturn} \\ 
        Phoebe & 120 & 8.30$\times10^{18 }$ & \cite{Williams_NSSDCA_SaturnianFact} \\ 
        Paaliaq & 11 & 4.30$\times10^{15 }$ & \cite{Denk_OuterMoonsSaturn} \\ 
        Skathi & 4 & 1.00$\times10^{14 }$ & \cite{Denk_OuterMoonsSaturn} \\ 
        Albiorix & 16 & 7.00$\times10^{15 }$ & \cite{Denk_OuterMoonsSaturn} \\ 
        Bebhionn & 3 & 5.00$\times10^{13 }$ & \cite{Denk_OuterMoonsSaturn} \\ 
        Erriapus & 5 & 3.00$\times10^{14 }$ & \cite{Denk_OuterMoonsSaturn} \\ 
        Siarnaq & 20 & 1.60$\times10^{16 }$ & \cite{Denk_OuterMoonsSaturn} \\ 
        Skoll & 3 & 3.00$\times10^{13 }$ & \cite{Denk_OuterMoonsSaturn} \\ 
        Tarvos & 7.5 & 8.00$\times10^{14 }$ & \cite{Denk_OuterMoonsSaturn} \\ 
        Tarqeq & 3.5 & 6.00$\times10^{13 }$ & \cite{Denk_OuterMoonsSaturn} \\ 
        Greip & 3 & 3.00$\times10^{13 }$ & \cite{Denk_OuterMoonsSaturn} \\ 
        Hyrrokkin & 4 & 1.00$\times10^{14 }$ & \cite{Denk_OuterMoonsSaturn} \\ 
        Mundilfari & 3.5 & 9.00$\times10^{13 }$ & \cite{Denk_OuterMoonsSaturn} \\ 
        Jarnsaxa & 3 & 2.00$\times10^{13 }$ & \cite{Denk_OuterMoonsSaturn} \\ 
        Narvi & 3.5 & 1.00$\times10^{14 }$ & \cite{Denk_OuterMoonsSaturn} \\ 
        Bergelmir & 3 & 3.00$\times10^{13 }$ & \cite{Denk_OuterMoonsSaturn} \\ 
        Suttungr & 3.5 & 9.00$\times10^{13 }$ & \cite{Denk_OuterMoonsSaturn} \\ 
        Hati & 3 & 3.00$\times10^{13 }$ & \cite{Denk_OuterMoonsSaturn} \\ 
        Bestla & 3.5 & 8.00$\times10^{13 }$ & \cite{Denk_OuterMoonsSaturn} \\ 
        Thrymr & 3.5 & 1.00$\times10^{14 }$ & \cite{Denk_OuterMoonsSaturn} \\ 
        Aegir & 3 & 2.00$\times10^{13 }$ & \cite{Denk_OuterMoonsSaturn} \\ 
        Kari & 3.5 & 6.00$\times10^{13 }$ & \cite{Denk_OuterMoonsSaturn} \\ 
        Fenrir & 2 & 1.00$\times10^{13 }$ & \cite{Denk_OuterMoonsSaturn} \\ 
        Surtur & 3 & 1.50$\times10^{13 }$ & \cite{Denk_OuterMoonsSaturn} \\ 
        Ymir & 9 & 2.00$\times10^{15 }$ & \cite{Denk_OuterMoonsSaturn} \\ 
        Loge & 3 & 3.00$\times10^{13 }$ & \cite{Denk_OuterMoonsSaturn} \\ 
        Fornjot & 3 & 5.00$\times10^{13 }$ & \cite{Denk_OuterMoonsSaturn} \\ \hline
    \end{tabular}
\end{table}
\clearpage

\section{Estimation of the impact of non-conservative forces}\label{app:nonconservative}
In the present study, we have neglected non-conservative forces such as the tides, and the effects arising from the solar radiation. In this appendix, we estimate their order of magnitude.
\subsection{Tidal forces}
The effect of tides on submoon survival can be estimated by computing the maximal survival time of a submoon, beyond which it would necessarily be ejected or collide with its host satellite. This quantity can be computed using Equation (12) from \cite{Barnes_2002}. 
\begin{equation}
    T_{\max }=\frac{2}{13}\left[\frac{\left(f a_{moon}\right)^3}{3 M_p}\right]^{13 / 6} \frac{M_{moon}^{8 / 3} Q_{moon}}{3 k_{2 moon} M_{sub} R_{moon}^5 \sqrt{G}}
\end{equation}

For submoons as small as those we consider in the present study ($M_{sub}\approx4.9\times 10^9$ kg), assuming a rather high $k_2/Q$ value of $1.2\times10^{-3}$ \citep{Rosario-Franco_2020}, and a small $f$ value of 0.33 \citep{Rosario-Franco_2020}, the migration time is of the order of magnitude of at least $10^{11}$ yrs, as shown in Table \ref{tab:temps_marees}

\begin{table}[!ht]
    \centering
    \caption{Maximum tidal migration time for a submoon of the present study to collide or be ejected from its parent satellite.}
    \begin{tabular}{ll}
    \hline
        Satellite & $T_{max}$ (yrs) \\ \hline
        Ganymede & $5.1\cdot10^{14}$ \\ 
        Callisto & $1.3\cdot10^{16}$ \\ 
        Rhea & $5.2\cdot10^{11}$ \\ 
        Titan  & $1.4\cdot10^{16}$ \\ 
        Iapetus & $9.1\cdot10^{16}$ \\ 
        Phoebe & $1.8\cdot10^{18}$ \\ \hline
    \end{tabular}
    \label{tab:temps_marees}
\end{table}

\subsection{Forces arising from solar radiation}
The main effects arising from solar radiation are direct solar-radiation-pressure, Poynting-Robertson drag, the Yarkovsky–O'Keefe–Radzievskii–Paddack (YORP) effect, the Yarkovsky effect, and the BYORP effect.

\cite{BURNS1979} showed that the ratio $\beta$ of the radiation pressure force to the solar gravity force can be written as 
\begin{equation}
    \beta = F_r/F_g = 5.7\times10^{-5}\frac{Q_{pr}}{\rho R}
\end{equation}
where $F_r$ and $F_g$ are the solar-radiation-pressure and gravitational forces acting on the body, respectively; $\rho$ and $R$ are its density and physical radius; and $Q_{pr}$ is the radiation pressure efficiency factor. The quantities in the above equation are expressed in cgs.
The authors noted that this ratio is independent of the distance from the Sun and deduced that the perturbed orbits remain conic sections, meaning that radiation pressure can be modeled as a small, non-dissipative modification of solar gravity. In our case, assuming $Q_{pr}=1$ (typical for macroscopic objects such as the submoon), the ratio of the solar-radiation-pressure force to the solar gravitational force is approximately $6\times10^{-9}$. This means that the perturbation induced by the solar pressure is 8 to 9 orders of magnitude smaller than the solar gravitational perturbation, and is therefore negligible in our simulations.

In contrast, Poynting-Robertson drag is a dissipative force. It is mainly effective for small particles and can be neglected for objects with radii as large as 100~m. Equation 55 from \cite{BURNS1979} gives the characteristic time of the Poynting-Robertson effect for an object in a planetocentric orbit. 
\begin{equation}
    9.3\times10^6 \frac{r^2\rho R}{Q_{pr}} \text{ years}
\end{equation}
where $r$ is the heliocentric distance of the object in astronomical units, while $\rho$ and $R$ are again expressed in cgs units.
In our case, the particle is in a satellite-centered orbit, but this expression can still be used to estimate the order of magnitude of the effect. For a submoon orbiting at approximately Jupiter's heliocentric distance, the characteristic timescale is of order $10^{12}$~yr, exceeding the age of the Solar System by more than two orders of magnitude. Over the few million years considered here, this effect can therefore safely be neglected.

The YORP effect vanishes for perfectly spherical objects \citep{Breiter_2008}. Because the submoons were modeled as perfectly spherical in our simulations, this effect was not included. The BYORP (binary YORP) effect also vanishes for symmetrical bodies such as spheres or ellipsoids \citep{Jacobson_2011}.

Finally, the binary Yarkovsky effect could affect the dynamical evolution of such submoons by modifying their semi-major axes. \cite{Zhou_2024} showed that the semi-major axis migration rate $\dot{a}_Y$ of a secondary in a binary asteroid system due to the binary Yarkovsky can be written  

\begin{equation}
    \dot{a}_Y= \frac{2f_Y\mathcal{F}}{n}
\end{equation}
with $f_Y$, $\mathcal{F}$ and $n$ respectively a dimensionless Yarkovsky coefficient, the nominal radiation pressure per unit mass, and the mean motion of the secondary.

\cite{Zhou_2024} showed that $f_Y$ vanishes for synchronous rotators and is typically of order $\approx10^{-3}$ for asynchronous rotators. For prograde secondaries, the binary Yarkovsky effect tends to drive the secondary toward the synchronous state by modifying its orbit, after which the effect becomes negligible. As we did not account for the rotation of either the submoons or their host satellites in our simulations, the binary Yarkovsky effect was not computed. However, by assuming $f_Y =10^{-3}$, one can estimate the orbital evolution which could be induced by the binary Yarkovsky. Assuming an albedo of $0.05$ and a submoon located halfway across its host satellite's Hill sphere, the estimated migration rates induced by the binary Yarkovsky effect are listed in Table \ref{tab:submoon_migration} for selected satellites.

\begin{table}[!ht]
    \centering
    \caption{Characteristic submoon migration rates and stability limits for selected parent satellites.}
    \label{tab:submoon_migration}
    \centering
    \setlength{\tabcolsep}{3.5pt}
    \renewcommand{\arraystretch}{1.15}
    \begin{tabular}{lccc}
    \hline\hline
    Parent &
    $\dot{a}_Y$ &
    $r_{R}$ &
    $R_{H}$ \\
     &
    (km\,yr$^{-1}$) &
    (km) &
    (km) \\
    \hline
    Ganymede & $1.457\times10^{-6}$ & 4136 & 29533 \\
    Callisto  & $3.076\times10^{-6}$ & 3718 & 46684 \\
    Rhea      & $2.96\times10^{-7}$ & 1033 & 5431 \\
    Titan     & $8.78\times10^{-7}$ & 4005 & 48803 \\
    Iapetus   & $3.973\times10^{-6}$ & 953 & 33832 \\
    Phoebe    & $2.6504\times10^{-5}$ & 158 & 20422 \\
    \hline
    \end{tabular}
\end{table}

\section{Expression of the force in a planetocentric frame with oblate bodies}\label{app:force}

Equation (13) from \cite{Lainey2004} gives the formulation of the acceleration $\mathbf{\ddot{r}}_i$ of a body $P_i$ in a planetocentric $\mathcal{N}+1$ body problem with fixed axes centered on the body $P_0$: 

\begin{equation}
\begin{aligned}
    \ddot{\mathbf{r}}_i &= G\left(m_0 +m_i\right) \left[-\frac{\mathbf{r_{i}}}{r_{i}^3} - \nabla_{i0}V_{\bar{0} \hat{i}} + \nabla_{i}V_{\bar{i} \hat{0}}\right] \\&+
    G\sum_{j=1,j\ne i}^\mathcal{N} m_j \left[ \frac{\mathbf{r_{ij}}}{r_{ij}^3} - \nabla_{ij}V_{\bar{j} \hat{i}} + \nabla_{ji}V_{\bar{i} \hat{j}}    -   \frac{\mathbf{r_{j}}}{r_{j}^3} + \nabla_{j}V_{\bar{j} \hat{0}} -\nabla_{j0}V_{\bar{0} \hat{j}}\right] 
\end{aligned}
\end{equation}

with $m_j$, $\mathbf{r}_j$ and $r_j$ the mass, position vector and norm of position vector of body $P_j$. $\mathbf{r}_{ij} = \mathbf{r}_j-\mathbf{r}_{i}$ is a vector, and ${r}_{ij}$ is its norm. $V_{ij}$ is the force function (opposite of the gravitational potential) of the body $P_j$ acting on $P_i$.

$\nabla_{ji}$ is the gradient operator such that
\begin{equation}
\nabla_{ji}f
= \begin{pmatrix}
\displaystyle \frac{\partial f}{\partial x_{ji}} \\[6pt]
\displaystyle \frac{\partial f}{\partial y_{ji}} \\[6pt]
\displaystyle \frac{\partial f}{\partial z_{ji}}
\end{pmatrix}
\end{equation}
with $x_{ji}$, $y_{ji}$ and $z_{ji}$ the coordinates of $P_i$ with respect to $P_j$.
As in \cite{Lainey2004} and \cite{Dahoumane_2025}, $V_{ij}$ can be split in two different terms: $V_{ij}= V_{\bar{ij}} +V_{\bar{i} \hat{j}}$, with $V_{\bar{ij}}$ representing the action of a point-mass $P_j$ acting on $P_i$, while $V_{\bar{i} \hat{j}}$ represents the effect of the non-spherical gravitational potential of $P_j$ on $P_i$.
While $V_{\bar{ij}}= 1/r_{ij}$, $V_{\bar{j} \hat{i}}$ can be itself split in two terms \citep{Lainey2004,Dahoumane_2025}: $V^{(1)}_{\bar{i} \hat{j}}$ is the force function arising from the zonal terms of $P_j$ gravity field, while $V^{(2)}_{\bar{i} \hat{j}}$ arises from the sectoral terms of $P_j$. In the present work, we do not use the sectoral gravity terms for any of the simulated bodies, so $V^{(2)}_{\bar{i} \hat{j}}$ is assumed to be equal to zero, implying that $V_{\bar{i} \hat{j}} =V^{(1)}_{\bar{i} \hat{j}}$.

\cite{Lainey2004} gave a general expression for $V^{(1)}_{\bar{i} \hat{j}}$
\begin{equation}
    V^{(1)}_{\bar{i} \hat{0}} = \sum^\infty_{n=2}-\frac{(R_0)^n}{r_i^{n+1}}J_nP_n(\sin{\phi_i)}
\end{equation}

with $R_0$, $J_n$, $P_n$ and $\phi_i$ respectively the equatorial radius of $P_0$, the $P_0$ zonal term of degree $n$, the Legendre polynomial of degree $n$, and the latitude of $P_i$ with respect to $P_0$.

By analogy with this expression, we can write the general expression:
\begin{equation}
    V_{\bar{i} \hat{j}}^{(1)}  = -\sum_{n=2}^{\infty} \frac{\left(R_j\right)^n}{r_{ji}^{n+1}}\left\{J_n^{(j)} P_n\left(\sin \phi_i^{(j)}\right)\right\}
\end{equation}

With $R_j$,$J_n^{(j)}$ and $\phi_i^{(j)}$ respectively the equatorial radius of body $P_j$, the zonal term of degree $n$ of $P_j$, and the latitude of $P_i$ with respect to the equator of $P_j$. For degree 2, we have: 

\begin{equation}
    V_{\bar{i} \hat{j}}^{(1)}  =  - \frac{R_j^2 J_2^{(j)}}{r_{ji}^{3}}\left(\frac{3}{2} \sin^2 \phi_i^{(j)}-\frac{1}{2}\right)
\end{equation}

Finally, we are now able to find the expression of $\nabla_{ji}V_{\bar{i} \hat{j}}$. We could also have found it by simply replacing the terms $R_0$, $\mathbf{k}$, $\mathbf{r}_i$ and $\sin\phi_i$ by $R_j$, $\mathbf{k}^{(j)}$, $\mathbf{r}_{ji}$ and $\sin \phi_i^{(j)}$, in equation (2) from \cite{Dahoumane_2025} , with $\mathbf{k}^{(j)}$ being the rotation axis of $P_j$. 

\begin{equation}
    \nabla_{ji}V_{\bar{i} \hat{j}}^{(1)} = \frac{3R_j^2 J_2^{(j)}}{r_{ji}^4}\left[\frac{1}{2}\frac{\mathbf{r_{ji}}}{r_{ji}}\left(5\sin^2\phi_i^{(j)}-1\right)-\mathbf{k^{(j)}}\sin\phi_i^{(j)}\right]
\end{equation}

with
\begin{equation}
    \sin\phi_i^{(j)} = \frac{\mathbf{r_{ji} \cdot k^{(j)}}}{r_{ji}}
\end{equation}
In the present case, as the only harmonic coefficients considered are the zonal ones, we have $\nabla_{ji}V_{\bar{i} \hat{j}}^{(1)} = \nabla_{ji}V_{\bar{i} \hat{j}}$

\section{Distribution of submoons surviving between 1 and 2 Myr}

\begin{figure*}[htbp]
  \centering

  \includegraphics[width=\linewidth]{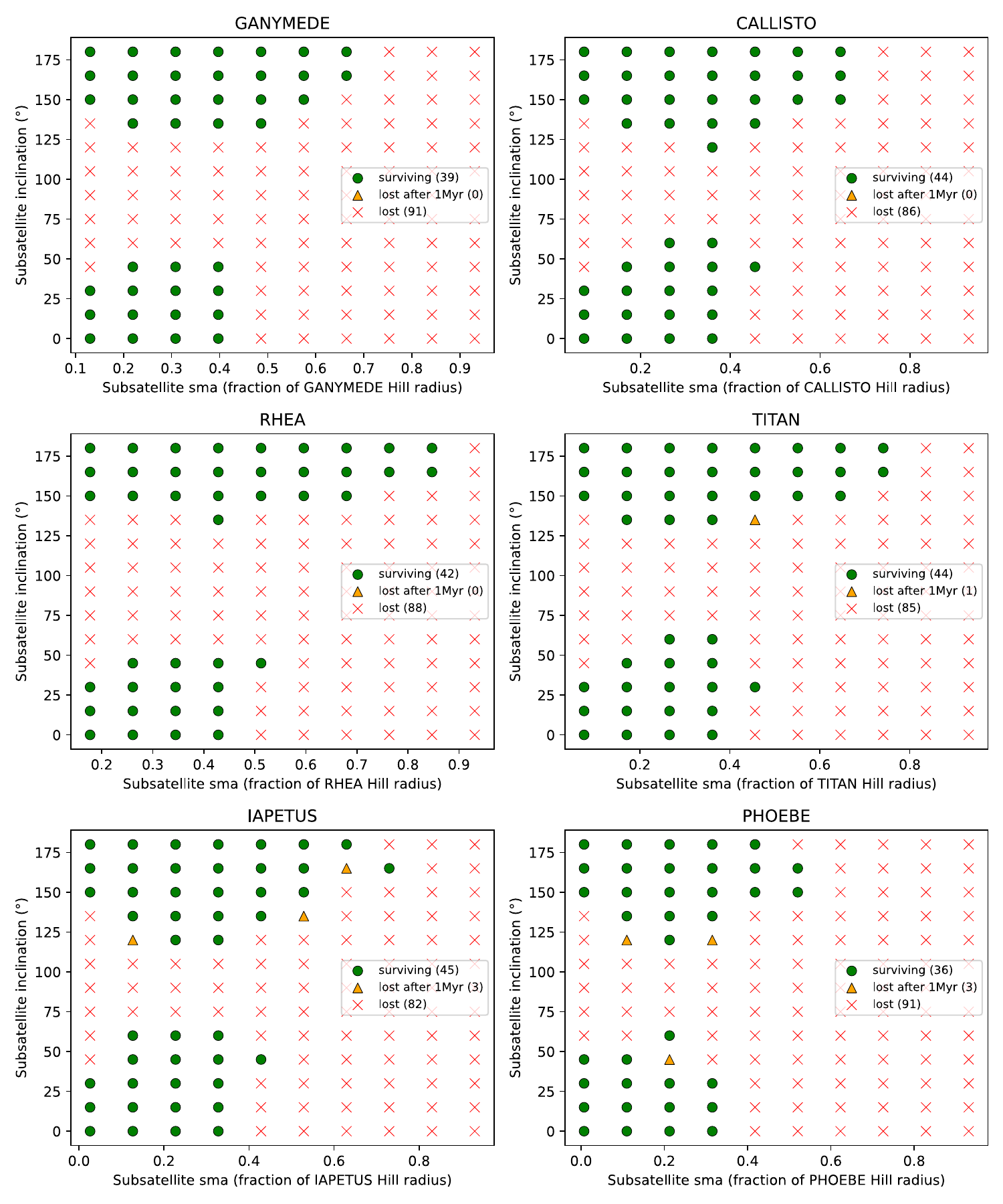}
  \caption{Distribution of submoon configurations around Ganymede, Callisto, Rhea, Titan, Iapetus, and Phoebe. The symbols distinguish submoons that survived to 2 Myr, those lost before 1 Myr, and those lost between 1 and 2 Myr. The inclinations are given with respect to the satellite orbital plane around the planet.}
  \label{fig:1_5myr}
\end{figure*}
\end{appendix}

\end{document}